\documentclass[aps,prd,preprint,nofootinbib,longbibliography]{revtex4-1}

\usepackage{amsmath}
\usepackage{amssymb} 
\usepackage[cm]{fullpage}
\usepackage{graphicx} 
\usepackage{siunitx}
\usepackage{tabularx} 
\usepackage[colorlinks,linkcolor=blue,citecolor=blue,urlcolor=blue]{hyperref}
\usepackage[table,xcdraw]{xcolor}
\usepackage[utf8]{inputenc}

\DeclareSIUnit\parsec{pc}

\newcommand{\beq}{\begin{equation}}
\newcommand{\eeq}{\end{equation}}
\newcommand{\beqx}{\begin{equation*}}
\newcommand{\eeqx}{\end{equation*}}

\newcommand{\eps}[1]{\epsilon_{#1}}
\newcommand{\del}[1]{\delta_{#1}}

\newcommand{\lnp}[1]{\ln{\left(#1\right)}}
\newcommand{\rpar}[1]{\left(#1\right)}
\newcommand{\spar}[1]{\left[#1\right]}

\newcommand{\bd}{{\rm d}}

\begin{document}

\title{Reheating in Gauss-Bonnet-coupled inflation}
\date{\today}
\author{Carsten van de Bruck}
\affiliation{Consortium for Fundamental Physics, School of Mathematics and Statistics, University of Sheffield, Hounsfield Road, Sheffield S3 7RH, United Kingdom}
\author{Konstantinos Dimopoulos}
\affiliation{Consortium for Fundamental Physics, Physics Department,Lancaster University, Lancaster LA1 4YB, United Kingdom}
\author{Chris Longden}
\affiliation{Consortium for Fundamental Physics, School of Mathematics and Statistics, University of Sheffield, Hounsfield Road, Sheffield S3 7RH, United Kingdom}

\begin{abstract}

We investigate the feasibility of models of inflation with a large Gauss-Bonnet coupling at late times, which have been shown to modify and prevent the end of inflation. Despite the potential of Gauss-Bonnet models in predicting favourable power spectra, capable of greatly lowering the tensor-to-scalar-ratio compared to now-disfavoured models of standard chaotic inflation, it is important to also understand in what context it is possible for post-inflationary (p)reheating to proceed and hence recover an acceptable late-time cosmology. We argue that in the previously-studied inverse power law coupling case, reheating cannot happen due to a lack of oscillatory solutions for the inflaton, and that neither instant preheating nor gravitational particle production would avoid this problem due to the persistence of the inflaton's energy density, even if it were to partially decay. Hence we proceed to define a minimal generalisation of the model which can permit perturbative reheating and study the consequences of this, including heavily modified dynamics during reheating and predictions of the power spectra.

\end{abstract}

\maketitle

\tableofcontents

\section{Introduction}

The inflationary paradigm, in which the accelerating expansion of the early universe is invoked to explain the flatness and homogeneity of the observable universe, as well as account for the nearly scale-invariant spectrum of primordial density fluctuations now measured by Planck \cite{Planck2015Inflation}, has been the dominant theory of early-universe cosmology for several decades. Despite its successes, however, the simplest models of chaotic inflation are now experimentally disfavoured due to predicting an overly large tensor-to-scalar ratio \cite{Ade:2015tva}. As such, extended models of inflation involving modified theories of gravity such as Higgs inflation \cite{Bezrukov:2007ep} and Starobinsky inflation \cite{Starobinsky:1980te} which can produce much smaller, and hence experimentally compatible, tensor amplitudes, are now attracting considerable interest. As the available experimental data becomes more and more precise, inflationary cosmology will continue to act as a stage on which such new physics in the early universe can be increasingly well studied.

Another such extended theory of inflation which has been studied is a scalar field coupled to the Gauss-Bonnet combination of quadratic curvature scalars ($R^2 - 4 R^{\mu \nu} R_{\mu \nu} + R^{\rho \mu \sigma \nu} R_{\rho \mu \sigma \nu}$) \cite{Guo:2010jr,Kawai:1998ab,Kawai:1999pw,Satoh:2008ck,Jiang:2013gza,Koh:2014bka}. Appearing in string theory, higher-dimensional Brane World models \cite{Binetruy:2002ck,Brax:2003fv,Charmousis:2002rc,Gasperini:1996fu,Germani:2002pt,Gravanis:2002wy,Tsujikawa:2004dm}, as a component of Horndenski theories \cite{Horndeski:1974wa,Kobayashi:2011nu} and their so-called `Fab Four' subset \cite{Charmousis:2011ea,Copeland:2012qf,Kaloper:2013vta}, the presence of the Gauss-Bonnet term is not uncommon in theoretical physics. Furthermore, the presence of higher-power curvature scalars in UV-complete theories of gravity is generally expected from a bottom-up effective field theory perspective \cite{Donoghue:2012zc}. The Gauss-Bonnet term has also been studied in the context of bouncing cosmologies \cite{Bamba:2014mya}. Given the presence of cosmological models including a Gauss-Bonnet term in the literature, whether used to realise inflation or otherwise, as well as its status as a simple and perhaps physically motivated correction to Einstein gravity, we argue that it is worthwhile to further our understanding of the cosmological implications of Gauss-Bonnet related phenomena, and what observational signals may allow us to confirm or exclude its presence in whatever theory of high-energy physics correctly describes the early universe.

In this work we are going to particularly focus on the case where the coupling function between the inflaton and the Gauss-Bonnet term is an inverse power law, following on from previous work such as \cite{Guo:2010jr}, which is interesting as a simple prototype of coupling functions which grow large as the inflaton field becomes small towards the end of inflation. While such previous slow-roll analyses showed that such models can produce interesting, experimentally-compatible, primordial power spectra at the time of horizon crossing, they did not account for the effects of the large Gauss-Bonnet coupling at the end of inflation, which only become apparent when one looks beyond the slow-roll approximation. Indeed, as found in previous work \cite{MatThesis, PhysRevD.93.063519}, such a coupling prevents inflation from ending and hence has implications for the transition between the inflationary universe and the radiation-dominated epoch that must follow in order to allow nucleosynthesis and the standard big-bang late-time cosmology to proceed in a realistic fashion. We hence investigate in this work whether post-inflationary reheating can proceed in inverse power law coupled Gauss-Bonnet inflation. 

The layout of this paper proceeds as follows: In section \ref{sec:model} we define the model, study its late-time behaviour and show that the simplest mechanisms of (p)reheating do not proceed due to modifications of the late-time inflationary dynamics, in particular a lack of oscillations around the minimum of the inflation's potential. Next, in section \ref{sec:nonpert}, we investigate the feasibility of non-standard reheating procedures which do not depend on oscillations, such as instant preheating, but find that even if radiation is produced by such a mechanism it cannot come to dominate the universe due to the persistence of the Gauss-Bonnet-coupled inflaton. In section \ref{sec:generalmodel} we present a minimalistic generalisation of our model in which standard perturbative reheating can proceed under certain conditions, but with modified dynamics, and investigate the implications of this in terms of links to experimental data in section \ref{sec:ResultsConstraints}. We conclude in section \ref{sec:Conclusion}.

\section{Gauss-Bonnet-coupled inflation} \label{sec:model}

We begin with the action describing a general relativity, plus a scalar field coupled to the Gauss-Bonnet term,

\beq \label{eq:GBaction}
S = \int \bd^4 x \sqrt{-g} \left[ \frac{1}{2} R - \frac{1}{2} \omega (\partial \phi)^2 - V(\phi) - \frac{1}{2} G(\phi) E + \mathcal{L}_\text{m} \right ] \, ,
\eeq
where $E = R^2 - 4 R^{\mu \nu} R_{\mu \nu} + R^{\rho \mu \sigma \nu} R_{\rho \mu \sigma \nu}$. The model is hence specified by two arbitrary functions, the potential $V(\phi)$ and the Gauss-Bonnet coupling $G(\phi)$. In general, previous studies of this model have included the constant $\omega = \pm 1$ to diversify the range of possible dynamics, but here we will set $\omega = 1$, limiting ourselves to scalar fields with canonical kinetic terms, in the absence of any particular physical motivation to study the $\omega = -1$ case. We also nominally include a matter Lagrangian as we are interested in reheating.

Variation of the action leads to the field equations,
\begin{align}
3 H^2 & = \frac{1}{2} \dot{\phi}^2 + V(\phi) +12 H^3 \dot{G} + \rho_\gamma \, , \label{eq:FriedmanGB1} \\
2 \dot{H} & = - \dot{\phi}^2 + 4 H^2 (\ddot{G}-H\dot{G}) + 8 H \dot{H} \dot{G} - \frac{4}{3} \rho_\gamma \,  ,
\end{align}
where we have assumed the matter sector of the model is a perfect relativistic fluid with $p_\gamma = \rho_\gamma / 3$. We also have the scalar equation of motion,

\beq \label{eq:GBKGE}
\ddot{\phi} + \rpar{3 H + \Gamma} \dot{\phi} + V_{,\phi} + 12 H^2 G_{,\phi} (\dot{H} + H^2) = 0  \, , 
\eeq
in which we have introduced the total decay rate into matter due to nongravitational quantum-corrections\footnote{We argue that because the effects being considered are limited to non-gravitational interactions, and we hence work essentially in flat spacetime when considering reheating, that the inclusion of the Gauss-Bonnet term does not modify the arguments leading to this standard description of perturbative reheating.}, $\Gamma$, in the usual way to model reheating \cite{Kofman:1997yn}. We obtain the value of $\Gamma$ using the standard tree-level result,

\beq
\Gamma = \frac{g^4 \sigma^2}{8 \pi m_\phi} + \frac{h^2 m_\phi}{8 \pi} \, ,
\eeq

for decays of the inflaton with mass $m_\phi$, to bosons with coupling constant $g^2 \sigma$ and to fermions with coupling constant $h$. We then assume typical coupling constants of, say, $\mathcal{O}(10^{-3})$ to specify the value of $\Gamma$ we will later use.

Accounting for decays, covariant conservation of stress-energy ($\nabla_\mu T^{\mu \nu}_{\phi} = - \nabla_\mu T^{\mu \nu}_{m} $) implies that the radiation fluid's energy density then evolves according to,
\beq\label{eq:rho_gamma}
\dot{\rho}_\gamma + 4 H \rho_\gamma = \Gamma \dot{\phi}^2  \, .
\eeq
Note that this equation is often seen instead in the form, $\dot{\rho}_\gamma + 4 H \rho_\gamma - \Gamma \rho_\phi = 0$, which implies $\dot{\rho}_\phi + 3 H \rpar{\rho_\phi + p_\phi} + \Gamma \rho_\phi = 0$ but this is inconsistent with eq. (\ref{eq:GBKGE}) as can be seen by using the definition of $\rho_\phi$ and $p_\phi$ from the Friedman equations. While the interaction term $\Gamma \rho_\phi$ is usable in standard chaotic inflation where $\rho_\phi = \dot{\phi}^2 / 2 + V$ and the inflaton oscillating about its minimum satisfies $V \approx \dot{\phi}^2 / 2$ such that $\rho_\phi \approx \dot{\phi}^2$, these approximations do not generally hold when a Gauss-Bonnet coupling is present.

We proceed to recursively define the usual slow-roll parameters $\eps{n}$,
\beq
\epsilon_0 = -\frac{\dot{H}}{H^2} \, , \quad \epsilon_n = \frac{\dot{\epsilon}_{n-1}}{H \epsilon_{n-1}} \, ,
\eeq
as well as the Gauss-Bonnet flow functions (described in e.g. \cite{Guo:2010jr, PhysRevD.93.063519}), $\del{n}$,
\beq
\delta_0 = 4 \dot{G} H \, , \quad \delta_n = \frac{\dot{\delta}_{n-1}}{H \delta_{n-1}} \, .
\eeq
At early times, as is usually the case with inflation, these parameters are small and obey $\eps{n} \, ,\del{n} \ll 1$, and it is in this regime that observable modes in the primordial power spectrum leave the horizon. An analysis based on these slow-roll parameters can hence be used to determine quantities of interest such as the tensor to scalar ratio, $r$, given some specific forms for the potential and coupling function. Following on from previous work, we are interested in the class of models where the potential is a positive power law and the coupling function is a negative power law, that is,

\beq \label{eq:basecouplings}
V(\phi) = V_0 \phi^n \, , \quad G(\phi) = G_0 \phi^{-m} \, ,
\eeq
and in particular, we will mostly discuss the case $n = m$ for simplicity, though many of the points discussed will hold for other choices, at least qualitatively. It is helpful to define the combination,

\beq
\alpha = \frac{4 V_0 G_0}{3} \, ,
\eeq
as an alternative parametrisation of the strength of Gauss-Bonnet coupling, as it is this combination which appears in many of the results. In particular a leading order slow-roll analysis shows that $r \propto (1 - \alpha)$ \cite{Guo:2010jr}. That is, increasing $\alpha$ reduces the tensor to scalar ratio, and the inclusion of a Gauss-Bonnet coupling can hence bring models with overly large tensor amplitudes back into agreement with experiment with a large enough $\alpha$. Using $\alpha$ is also convenient from the perspective that we can impose $\alpha \leq 1$, as above this limit, the Gauss-Bonnet term pushes the inflaton up its potential, leading to solutions in which $\phi$ grows, which we are not interested in.

\subsection{Modifications to late-time behaviour}

While the above slow-roll analysis reveals the intriguing ability to build inflationary models with spectral properties in agreement with experiment using a Gauss-Bonnet coupling, the first focus of this work is to show that it is not sufficient to simply read off these slow-roll predictions and declare the model a feasible explanation of what we know about inflationary dynamics. As discussed previously in \cite{PhysRevD.93.063519}, the inverse power law form of $G$ affects the end of inflation. As the field approaches its minimum at $\phi = 0$, usually it will pass through the minimum and begin to oscillate, however in this model the inflaton is inhibited from undergoing this usual late-time behaviour because as $\phi \rightarrow 0$, $G \rightarrow \infty$. 

\begin{figure}[t]
    \centering
    \includegraphics[width=\textwidth]{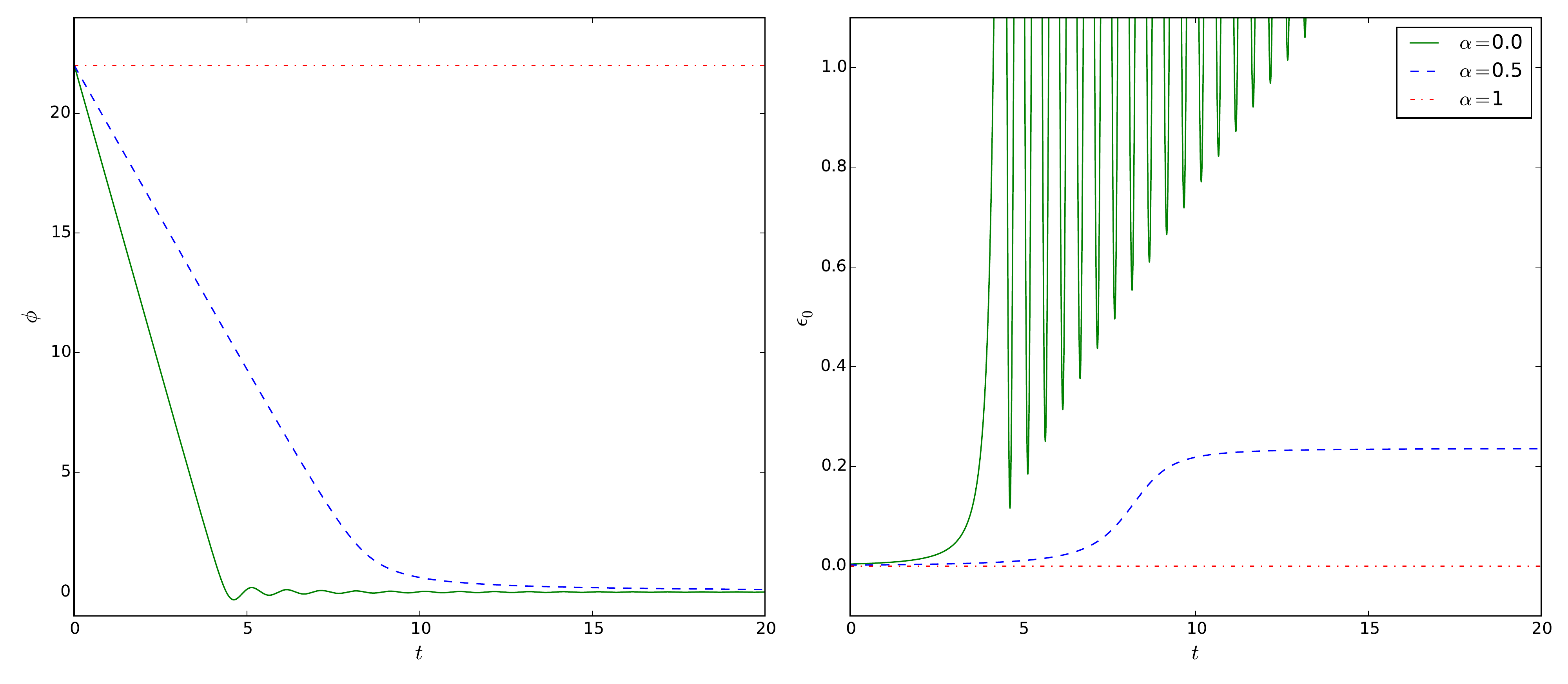}
    \caption{$\phi$ (left panel) and $\epsilon_0$ (right panel) for three different values of $\alpha$, in the model with $n=m=2$, with $V_0 = m_\phi^2 = (5.34 \times 10^{-6})^2$ and time shown in units of $m_\phi / 2 \pi$. In the first case, represented by the solid green line, $\alpha = 0$ and we have standard inflation, in which the field slow-rolls down to its minimum, oscillates around it, and inflation ends ($\epsilon_0 > 1$), allowing reheating to proceed and the standard late-time cosmology to be recovered. However, for the second case, represented by a dashed blue line, $\alpha = 0.5$ and as discussed in the main text, $\epsilon_0$ approaches a constant value, and $\phi$ asymptotically approaches $0$. $\phi$ is not allowed to decay significantly, and the continuously accelerating expansion of spacetime dilutes any matter that is produced to negligible levels. Inflation continues forever and we cannot make contact with standard late-time cosmology and recover the successes of the hot big bang model. Lastly, we also show the case of $\alpha = 1$. This is the dotted red line on the plots above, in which $\phi$ is constant and $\epsilon_0$ remains $0$ forever; a perfect de-Sitter expansion. }
    \label{fig:reheatingf1}
\end{figure}

In this regime, $\epsilon_0$ is observed numerically to approach a constant value and less than $1$, such that inflation does not end. The constant value $\epsilon_{0}$ approaches depends non-trivially on the value of $\alpha$, but it is always between $0$ (for $\alpha = 1$) and $1$ (for very small $\alpha$, $\epsilon$ will almost reach $1$). Similarly, we find that if $\epsilon_{0}$ is constant, $\delta_{0}$ must also be a constant between $0$ and $1$, which can be shown using the field equations to be given by,

\beq \label{eq:epsdeltalink}
\delta_0 \rpar{t \rightarrow \infty} = \frac{2 \epsilon_0}{1 + \epsilon_0} \, .
\eeq
We can also use the fact that $\epsilon_{0}$ is approximately a constant at late times to determine how $H$ evolves in time in this regime, by solving,

\beq \label{eq:Hevolutionconstanteps}
\dot{H} = -\epsilon_0 H^2 \, \quad \Rightarrow \quad H(t) = \rpar{c +  \epsilon_0 t}^{-1} \, ,
\eeq
where $c$ is a constant of integration. Knowing how $H$ evolves, we can infer the time-dependence of of $\phi$ by using eq. (\ref{eq:FriedmanGB1}), which, in terms of slow-roll parameters can be written,

\beq
V(\phi) =   \spar{3 - \eps{0} - \frac{1}{2} \delta_0 \rpar{5 + \delta_{0} - \eps{0}}} H^2 = \beta H^2 \, .
\eeq
That is, $V$ and $H^2$ are in direct proportion, and the combination of slow-roll parameters that serves as the constant of proportionality, $\beta$, between them is taken to be constant, as discussed above. Given that $V = V_0 \phi^n = \beta H^2$ and using the behaviour of $H$ determined in eq. (\ref{eq:Hevolutionconstanteps}), we can show that in the late-time regime when slow-roll parameters are constant,

\beq \label{eq:latetimephi}
\phi(t)=  \rpar{\frac{\beta n}{V_0}}^{\frac{1}{n}}\rpar{c +  \epsilon_0 t}^{-\frac{2}{n}} \, .
\eeq
Hence at late times, $\phi$ asymptotically approaches $0$, rather than oscillating about its potential minimum. This can also be solved this in terms of the e-folding number $N$ as a time coordinate to find that $\phi \propto e^{-2 \epsilon_0 N / n}$. We conclude that the inverse power law Gauss-Bonnet coupling overdamps the motion of the field $\phi$ by becoming large as $\phi \rightarrow 0$. The inflaton experiences a greater impedance from its Gauss-Bonnet coupling the closer it approaches its potential minimum, hence exponentially slowing its approach to that point. Note that this result contradicts the leading order slow-roll analysis (see \cite{Guo:2010jr}), which gives $\epsilon_0 \propto \phi^{-2}$ and would hence allow inflation to end as at small enough $\phi$, we could have $\epsilon_0 > 1$, and reheating would be expected to proceed normally after this. What we see now, looking beyond the slow-roll approximation, however, is that inflation will not end, the field will not oscillate and standard perturbative reheating will not occur. Figure \ref{fig:reheatingf1} demonstrates this using numerical solutions of the equations of motion including the decay term. We further note that, of course, in the absence of late-time oscillations in the inflaton, it is not only perturbative reheating, but also the resulting onset of parametric resonance required for non-perturbative preheating that fails. 

A particular case of this behaviour of interest is that when $\epsilon_0 = 0$, according to eq. (\ref{eq:latetimephi}), $\phi$ is a constant. Looking for constant solutions of eq. (\ref{eq:GBKGE}) by setting $\dot{\phi} = \ddot{\phi} = \dot{H} = 0$ and using the field equations, we find the condition for a solution to exist for the case where $n=m$ is that $\alpha = 1$, which is also demonstrated in figure \ref{fig:reheatingf1}.

\begin{figure}[t]
    \centering
    \includegraphics[width=\textwidth]{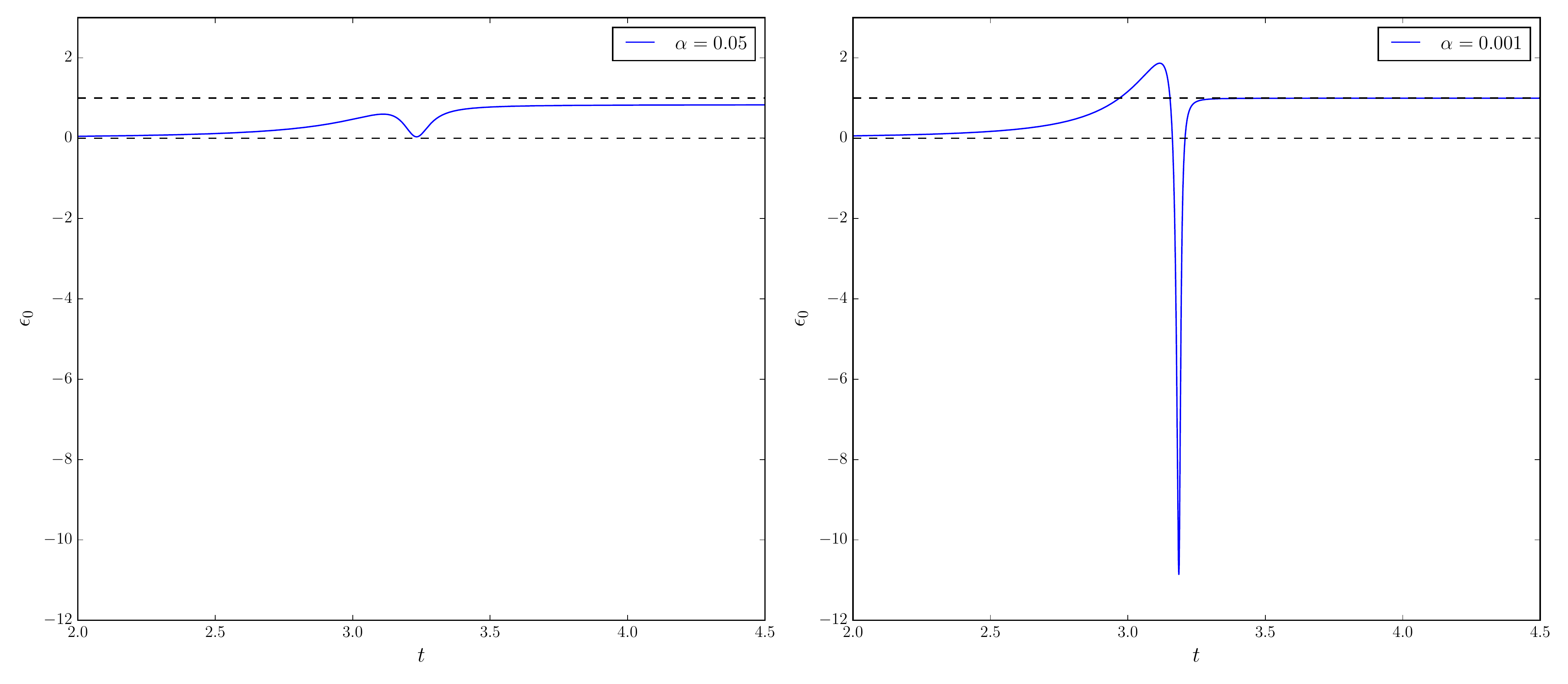}
    \caption{Two trajectories for $\epsilon_0$ with especially small values of $\alpha$ (0.05 on the left, and 0.001 on the right). Again, $n = m = 2$, $V_0 = m_\phi^2 = (5.0 \times 10^{-6})^2$ and time is shown in units of $m_\phi / 2 \pi$. Here we note that the transition from $\epsilon_0 \approx 0$ to its late-time constant value is no longer monotonous and can vary quite extremely for a very short amount of time of $\mathcal{O}(m_\phi / 2 \pi)$. The black dashed lines on these plots represent $\epsilon_0 = 0$ and $1$ respectively.}
    \label{fig:reheatingf2}
\end{figure}

As we saw in figure \ref{fig:reheatingf1}, following the initial period of $\epsilon_0 \approx 0 $, there is a step-like transition to the late-time constant value of $\epsilon_0$. For particularly small values of $\alpha$, we find this transition is no longer monotonic, and involves a sharp feature in which $\epsilon_0$ may even briefly fall outside the range $[0,1]$. Examples of this are shown in figure \ref{fig:reheatingf2}. While we find these features alone are incapable of facilitating reheating, we find it interesting that such a coupling to the Gauss-Bonnet term can create a localised feature in time, in which the expansion rate of space is violently changing -- we observe in particular that these transitions are accompanied by a sharp feature in $H$. This may be useful in increasing the efficiency of processes such as gravitational particle production.

While it is feasible that we could solve the problems presented here by introducing additional fields which account for reheating without affecting the late-time behaviour of the Gauss-Bonnet-coupled scalar $\phi$, we will not focus on this possibility in this paper. While extended theories of physics may predict large numbers of scalar fields which could be exploited for this purpose, we wish to first investigate approaches which do not require us to appeal to this. Nevertheless, it is potentially of some interest because if reheating could be made to proceed via a second field without changing the late-time behaviour of $\phi$, we could have a realisation of so-called quintessential inflation \cite{Dimopoulos:2001ix,Hossain:2014zma} in which the Gauss-Bonnet-coupled field serves as dark energy in the present epoch.

\section{Instant preheating and gravitational particle production} \label{sec:nonpert}

A mechanism that has been used to produce particles in non-oscillating theories is instant preheating \cite{Felder:1998vq, Hossain:2014xha}. Typically this occurs when after inflation, instead of oscillating around a local minimum, the field $\phi$ continues to grow in magnitude. This would happen, for example, when the potential is flat or steeply decreasing after inflation. Then, through a coupling to matter fields of the form, $\mathcal{L} = - \frac{1}{2} g^2 \phi^2 \chi^2$ the effective mass of $\chi$ particles, $g | \phi |$, would grow along with the magnitude of the inflaton. With this, even if only a small number of $\chi$ particles are produced by inflaton decays due to the lack of an oscillatory period, under the right conditions, the decay products can still come to dominate the universe due to their increasing mass. Unfortunately in our case, this mechanism is also inhibited because at late times we do not have $\phi$ growing, but rather an asymptotically approaching $0$. However, this could be made feasible again if we allow a shift in the coupling between $\phi$ and $\chi$. That is, $\mathcal{L} = - \frac{1}{2} g^2 (\phi - \nu)^2 \chi^2$, which could be motivated for example in the context of supersymmetric theories with superpotentials of the form $W = g \chi^2 (\phi - \nu)$ \cite{Felder:1998vq,Felder:1999pv,Berera:1998cq} or in the so-called A-term inflation scenario via an enhanced symmetry point of the potential \cite{Allahverdi:2006iq,Lyth:2006ec,BuenoSanchez:2006rze,Enqvist:2010vd}. With a coupling of this form, instant preheating would begin at the point $\phi = \nu$ and as in our model, $\phi$ then asymptotically approaches $0$, the produced $\chi$ particles could increase in effective mass by an amount $g \nu$.

We follow the standard approach to instant preheating as in \cite{Felder:1998vq}. Particle production occurs when the non-adiabaticity condition $|\dot{m_\chi}| > m_\chi^2$, or $g |\dot{\phi}| > g^2 (\phi - \nu)^2$. This implies that preheating takes place in a small region between the two points $\phi = \nu \pm \phi_*$ where $\phi_* = \sqrt{|\dot{\phi}|_{\phi = \nu}/g}$. In turn we infer that the time scale over which particles are produced is $\delta t = \phi_* / |\dot{\phi}|_{\phi = \nu}$, which is small (hence why this mechanism is known as instant preheating) and by the uncertainty principle the particles created would have approximate momenta of $k \sim 1/\Delta t = \sqrt{g|\dot{\phi}|_{\phi = \nu}}$. Hence as the field passes through the point $\nu$, following \cite{Felder:1998vq,Kofman:1997yn}, we expect the occupation number of the Fourier modes of $\chi$ to sharply increase such that,

\beq
n_k = \exp\rpar{-\frac{\pi (k^2 + m_\chi^2) }{g |\dot{\phi}|_{\phi = \nu}}} \, ,
\eeq
which implies a total number density of $\chi$ particles,

\beq
n_\chi = \frac{1}{2 \pi^2} \int^\infty_0 k^2 n_k \bd k  = \frac{(g |\dot{\phi}|_{\phi = \nu})^{3/2}}{8 \pi^3}\exp\rpar{-\frac{\pi m_\chi^2}{g |\dot{\phi}|_{\phi = \nu}}} \, .
\eeq
Assuming $\chi$'s bare mass is small, particularly compared to the momentum scale $\sqrt{g|\dot{\phi}|_{\phi = \nu}}$, the number density of particles is not exponentially suppressed, and the energy density of decay products following their production (so that $\phi < \nu$ and $m_\chi = |\phi - \nu| = \nu - \phi$) is given by,

\beq
\rho_\chi = m_\chi n_\chi = g (\nu - \phi) n_\chi = \frac{g^{5/2}|\dot{\phi}|_{\phi = \nu}^{3/2} (\nu - \phi)}{8 \pi^3} \, .
\eeq
Then, at late times, as $\phi \rightarrow 0$ according to eq. (\ref{eq:latetimephi}), this approaches the upper limit,

\beq
\rho_\chi \rightarrow \frac{g^{5/2}|\dot{\phi}|_{\phi = \nu}^{3/2} \nu}{8 \pi^3} \, .
\eeq
Returning to our investigations of the late-time behaviour of the Gauss-Bonnet coupled inflaton, we find even for somewhat optimistic parameters $g = 10^{-2}$, $\nu = 1$, $\alpha = 0.5$, that $|\dot{\phi}|_{\phi = \nu} \approx 5 \times 10^{-7}$, such that $\rho_\chi$ grows to a maximum of $\approx 1.5 \times 10^{-17}$, while at late times $\rho_\phi = \mathcal{O}(10^{-14})$. Usually, this would then proceed such that $\rho_\chi$, being composed of massive particles ($m_\chi \sim g \nu = 10^{-2} M_\text{Pl}$), would scale in time as $a^{-3}$, and the inflaton would either oscillate (e.g. in a $\phi^4$ potential, so $\rho_\phi \propto a^{-4}$) or enter a period of kinetic energy domination ($a^{-6}$) where the rate at which $\rho_\phi$ decreases with the scale factor is steeper than that of the massive decay products, and eventually the universe would be dominated by those decay products. We cannot, however, realise this in our model. As the effect of the Gauss-Bonnet coupling is to drive $\epsilon_0$ into an approximately constant regime between $0$ and $1$ at late times, the rate of energy loss of the inflation is $a^{-2\epsilon_0}$, and hence will not be susceptible to being outlasted by the decay products as in the usual situation with instant preheating. Given large enough values of $g$ and $\nu$, one could set the initial energy density of $\chi$ particles produced (before the expansion of the universe becomes important) to be larger than $\rho_\phi$ so that the universe would be briefly dominated by $\chi$, but this would require a coupling constant somewhat larger than the Planck mass, which in itself is undesirable. We expect a similar limitation to impede the progress of gravitational particle production, even in cases such as the trajectories in figure \ref{fig:reheatingf2} where the sharp changes in the expansion rate of the universe may give rise to more efficient than usual gravitational creation of particles.

\section{Perturbative reheating with generalised couplings} \label{sec:generalmodel}

One way of proceeding with this problem we found was to allow slightly more general couplings than those in eq. (\ref{eq:basecouplings}). In particular, as there is no special reason to impose that the bare vacuum expectation value of the inflaton potential coincides with the divergent point in the Gauss-Bonnet coupling, we consider functions of the form,

\beq \label{eq:sigmacouplings}
V(\phi) = V_0 (\phi+\varsigma)^n \, , \quad G(\phi) = G_0 \phi^{-n} \, .
\eeq
With the inclusion of a non-zero minimum in the potential at $\phi = \varsigma$,  we can potentially avoid the problem of the late-time damping of the inflaton as now it is free to cross and oscillate about the point $\varsigma$ due to a weaker Gauss-Bonnet coupling at this point, for $\varsigma$ sufficiently far from $0$. For convenience, we shall perform the field redefinition $\phi \rightarrow \phi - \varsigma$, and obtain,

\beq \label{eq:redefsigmacouplings}
V(\phi) = V_0 \phi^n \, , \quad G(\phi) = G_0 (\phi-\varsigma)^{-n} \, ,
\eeq
so that the new parameter $\varsigma$ is considered a parameter of the interaction between the Gauss-Bonnet term and the inflaton. This new model can also be seen as a prototype for other choices of the coupling which become largest (but not necessarily infinite so as in the pure inverse-power-law case) as the inflaton reaches its minimum, should some other theory or argument motivate such a choice of coupling. 

\begin{figure}[t]
    \centering
    \includegraphics[width=\textwidth]{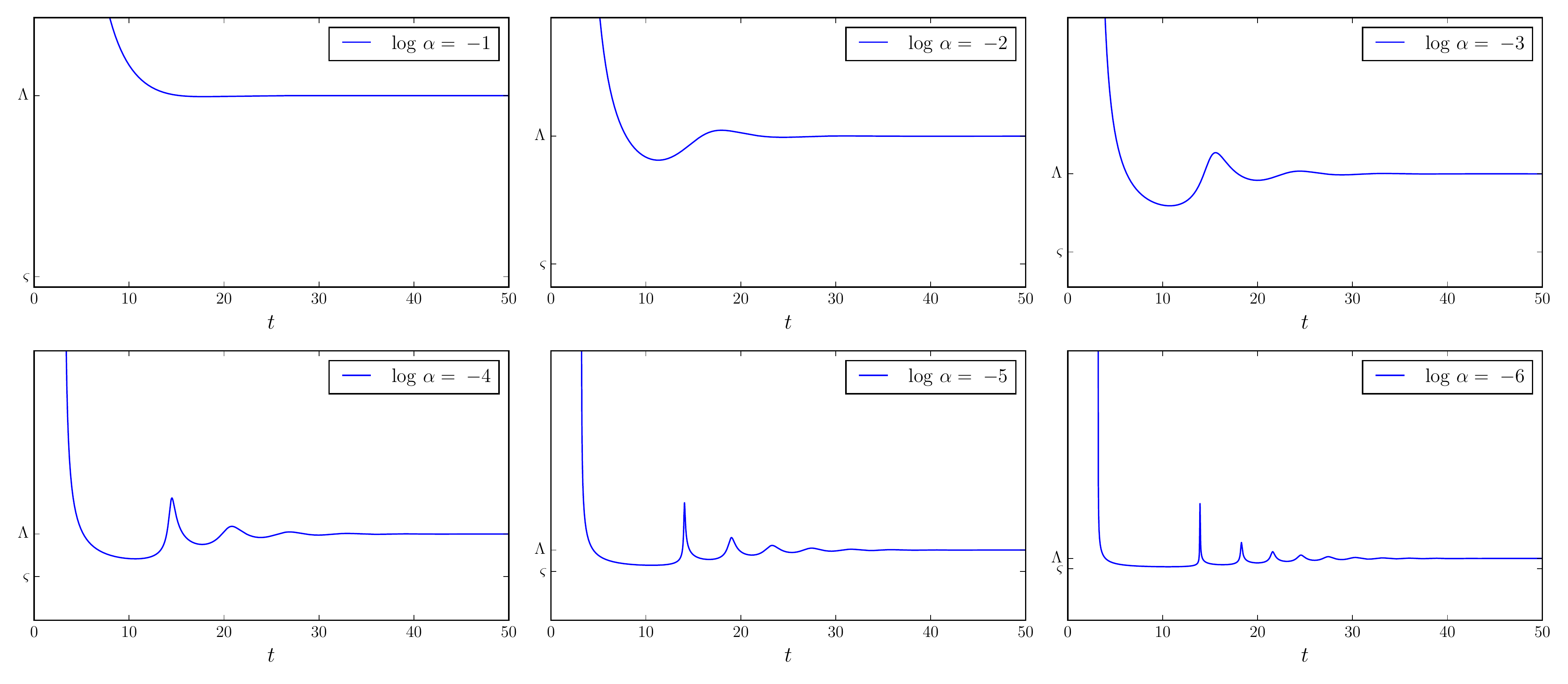}
    \caption{Six late-time evolutions of $\phi$ for the case $\varsigma = 0.05$ and $\alpha$ between $10^{-1}$ and $10^{-6}$. As $\alpha$ decreases (from top left to bottom right, along the rows) the strength of the Gauss-Bonnet coupling is decreasing and hence the damping effect becomes less extreme, allowing more oscillations to occur for a longer time. Oscillations in each case occur about the point $\Lambda$ as defined by eq. (\ref{eq:GBLambdagen}). The oscillations are non-sinusoidal, with the deformation in the waveform due to the varying strength of the Gauss-Bonnet coupling. In particular, as $\phi$ approaches the point $\varsigma$ where the coupling is infinite, it experiences a larger force returning it to the equilibrium at $\Lambda$ than it experiences for $\phi > \Lambda$ where the Gauss-Bonnet coupling is small. Time is shown in units of $m_\phi / 2 \pi$, and $n = m = 2$. }
    \label{fig:reheatingf3}
\end{figure}

Depending on the magnitude of $\varsigma$ there are multiple regimes of distinct phenomenology. Firstly, for positive $\varsigma$, the Gauss-Bonnet coupling diverges at positive values of $\phi$. Assuming that we consider model of inflation where the field $\phi$ begins at large positive values, this leads to a similar problem to the $\varsigma$ = 0 case in that the inflaton is impeded from rolling down its potential before even approaching the minimum; instead of asymptotically approaching zero, the field will approach a different  constant value that we shall call $\Lambda$, whose value we find (by solving the equation of motion for a constant field) to be,

\beq \label{eq:GBLambdagen}
\Lambda = \frac{\varsigma}{1 - \alpha^\frac{1}{n+1}} \, , \quad (\alpha \neq 0 \, ,\varsigma > 0) \, .
\eeq 
Indeed, even if we choose parameters such that the constant value $\phi$ should approach is larger than the initial condition for $\phi$, we have observed numerically that the field will increase to approach this value. Since, however, $\Lambda \neq \varsigma$, and the Gauss-Bonnet coupling is still finite at this point, it is possible that oscillations about this point could occur. Numerically we study this possibility for a range of $\alpha$ values and find that for very small $\alpha$, it is possible for very small, very short-lived and highly damped non-sinusoidal oscillations to occur about the point $\Lambda$. A series of plots of this are shown in figure \ref{fig:reheatingf3}.

Despite oscillatory late-time behaviour being possible in this regime, we find perturbative reheating cannot proceed nevertheless. The oscillations are too brief and low-amplitude to kick-start a significant enough amount of particle production to lead to radiation domination. Furthermore, as the field behaves like a cosmological constant, the expansion of the universe settles into a de-Sitter phase with only small deviations from exponential growth due to the minute oscillations, and any small amount of radiation that manages to be produced is quickly diluted away.

\begin{figure}[!ht]
    \centering
    \includegraphics[width=\textwidth]{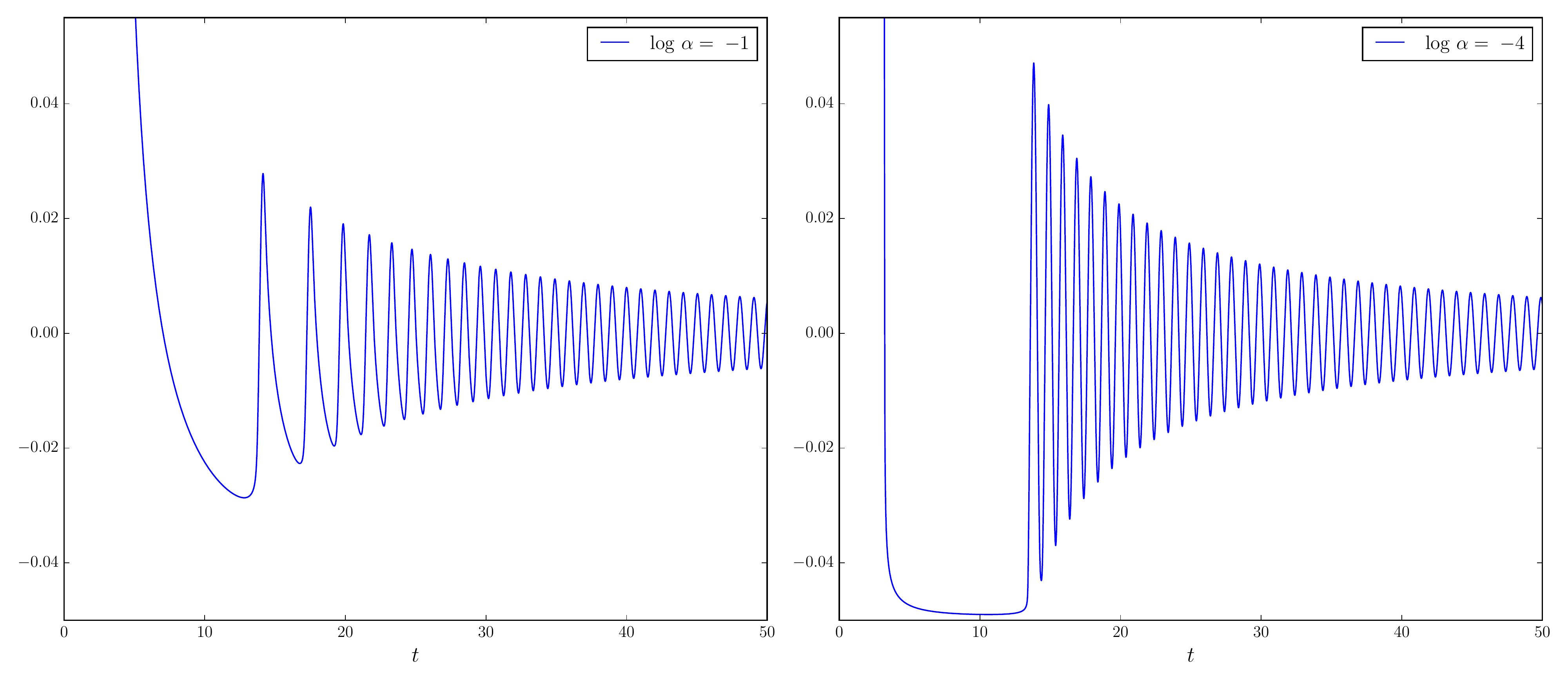}
    \caption{Two late-time evolutions of $\phi$ with $\varsigma = -0.05$ and $\alpha = 10^{-1}$ (left) or $10^{-4}$ (right). Non-sinusoidal oscillations with smaller amplitudes and lower frequencies than usual occur about $\phi = 0$. As the oscillations become smaller due to the expansion of the universe, the oscillations approach a more conventional waveform as the Gauss-Bonnet coupling grows weaker. Larger values of $\alpha$ unsurprisingly lead to smaller amplitudes and behaviour generally further away from the standard case. Time is again shown in units of $m_\phi / 2 \pi$, and $n = m = 2$.  }
    \label{fig:reheatingf4}
\end{figure}

To proceed further we must consider negative values of $\varsigma$. This is a more hopeful approach in the first place, as the divergence in the Gauss-Bonnet coupling now occurs at negative field values and the field may roll down to its minimum where the expansion of the universe may be able to cease accelerating and give way to a radiation-dominated epoch. For large (negative) enough choices of $|\varsigma| \gg |\phi|$, in fact, the Gauss-Bonnet coupling is essentially constant and the equations of motion are well approximated by those of standard inflation. While this certainly avoids the problem of inflation not ending, it is comparable to not having a Gauss-Bonnet coupling at all and hence will not help us understand the problem. As in conventional reheating the inflaton undergoes oscillations of amplitude $\mathcal{O}(10^{-1})$, we expect $\varsigma$ to need to be of around this magnitude to significantly affect the late-time dynamics of the inflaton and give us some idea of what a non-trivial Gauss-Bonnet coupling may do during reheating. Based on this approach we present examples of the oscillations of the coupled inflaton in figure \ref{fig:reheatingf4}.

We see that while at first, the oscillations are highly non-standard, eventually the field's oscillations decrease in amplitude until they are sufficiently close to zero that the shifted Gauss-Bonnet coupling becomes weak ($|\phi| \ll |\varsigma|$). Following this, fairly standard oscillations proceed and reheating can successfully occur. However, the dynamics of reheating will be modified compared to the standard case, even once the oscillations have become approximately sinusoidal, and the Gauss-Bonnet coupling is hence not strongly affecting the inflaton's evolution. To see this, note that while the effect of the Gauss-Bonnet coupling on the Klein-Gordon equation (\ref{eq:GBKGE}) is $\mathcal{O}(H^4)$, its contribution to the Friedman equation (\ref{eq:FriedmanGB1}) is only $\mathcal{O}(H^3)$. As reheating occurs around timescales when $H \approx \Gamma$, which is small, the Gauss-Bonnet coupling will become irrelevant to the scalar field dynamics earlier than it ceases to have a significant effect on the expansion of the universe. We will investigate in detail the implications of these effects in the next section. However, as the initial oscillations are made to be smaller in amplitude and lower in frequency than usual by the coupling to the Gauss-Bonnet term, we qualitatively expect reheating to take longer to complete as radiation is produced more slowly \footnote{This is because the right-hand side of Eq. (\ref{eq:rho_gamma}) depends not only on the decay rate $\Gamma$ but also on the amplitude and frequency of the oscillations.}, and that the temperature at the end of reheating will be correspondingly cooler.

Finally it is noted that the bound  ($0 \leq \alpha \leq 1$) we found to  be necessary to get interesting inflationary behaviour in the standard case is no longer exact once the
$\varsigma$ parameter is accounted for. We find that a decent approximation for the new upper limit of viable $\alpha$ values at which the system will behave as a cosmological constant is,

\beq
\alpha_\text{max} \approx \rpar{1 - \frac{\varsigma}{\phi_0}}^{n+1} \, ,
\eeq
where $\phi_0$ is the initial condition for $\phi$ at the beginning of inflation. As this is typically large, and we are interested in small $\varsigma$ primarily, however, this is usually a very small effect and we will continue to approximate $\alpha = 1$ as the upper bound.

\section{Results and Constraints} \label{sec:ResultsConstraints}

To extract useful information on the effects of non-standard reheating due to a Gauss-Bonnet coupling, we need to link the dynamics of reheating to some useful quantity. We choose to hence look at how reheating affects the primordial power spectra of inflation. This is encoded in the well-known relation (found in e.g. \cite{Planck2015Inflation})

\beq \label{eq:krange_phys}
N_* \approx 67 - \lnp{\frac{k_*}{a_0 H_0}} + \frac{1}{4} \lnp{\frac{V_*^2}{\rho_\text{end}}} + \frac{1 - 3 w_\text{int}}{12(1 + w_\text{int})}\lnp{\frac{\rho_\text{th}}{\rho_\text{end}}} - \frac{1}{12}\lnp{g_\text{th}} \, ,
\eeq
where $N_*$ is the number of e-folds before the end of inflation at which the observable wavenumber today, $k_*$ is leaving the horizon, $V_*$ is the potential energy at this time, $a_0$ and $H_0$ are the scale factor and Hubble parameter today, $\rho_\text{end}$ is the energy density at the end of inflation, $\rho_\text{th}$ is that when the universe is thermalised (e.g. the end of reheating, which we will define more precisely below), and $g_\text{th} \approx 10^3$ is the number of relativistic degrees of freedom at this time. Finally, $w_\text{int}$ is the average equation of state of the universe's matter content during reheating, defined as:

\beq \label{eq:wintdef}
w_\text{int} = \frac{1}{N_\text{th} - N_\text{end}} \int^{N_\text{th}}_{N_\text{end}} w(N) \, \bd N \, .
\eeq
where $w = p / \rho$ is the equation of state.

Using eq. (\ref{eq:krange_phys}) we can quantify how much the observable window during inflation changes in position due to the effects of the reheating period. That is, how many e-folds before the end of inflation the observable scales today exited the horizon, and hence precisely when we should extract our spectral predictions from our model of inflation. Then, in principle, with precise enough data, one could distinguish two models of inflation with the same primordial power spectra as inflation ends, but different reheating dynamics, by considering the shift in the observable window. We will call this quantity $\Delta N$, defined by the term in eq. (\ref{eq:krange_phys}) pertaining to the effects of reheating, that is,

\beq
\Delta N = \frac{1 - 3 w_\text{int}}{12(1+w_\text{int})} \ln\rpar{\frac{\rho_\text{th}}{\rho_\text{end}}} \, .
\eeq

To compute this we hence need to know the averaged equation of state defined in eq. (\ref{eq:wintdef}), but as the equation of state can rapidly oscillate between fairly large values in our model, evaluating this numerically is somewhat challenging (as has also been pointed out in e.g. \cite{Martin:2010kz}). Instead, we find and use the analytical result, which is derived in the appendix,

\beq \label{eq:wintexact}
w_\text{int} = \frac{2}{3} \frac{\ln\rpar{H_\text{end} / H_\text{th}}}{\ln\rpar{a_\text{th} / a_\text{end}}} - 1~, 
\eeq

This expression, advantageously, only requires one to find the values of $a$ and $H$ at the end of inflation and at the end of reheating, rather than using information from the entire period of reheating. 

\begin{figure}[!ht]
    \centering
    \includegraphics[width=0.45\textwidth]{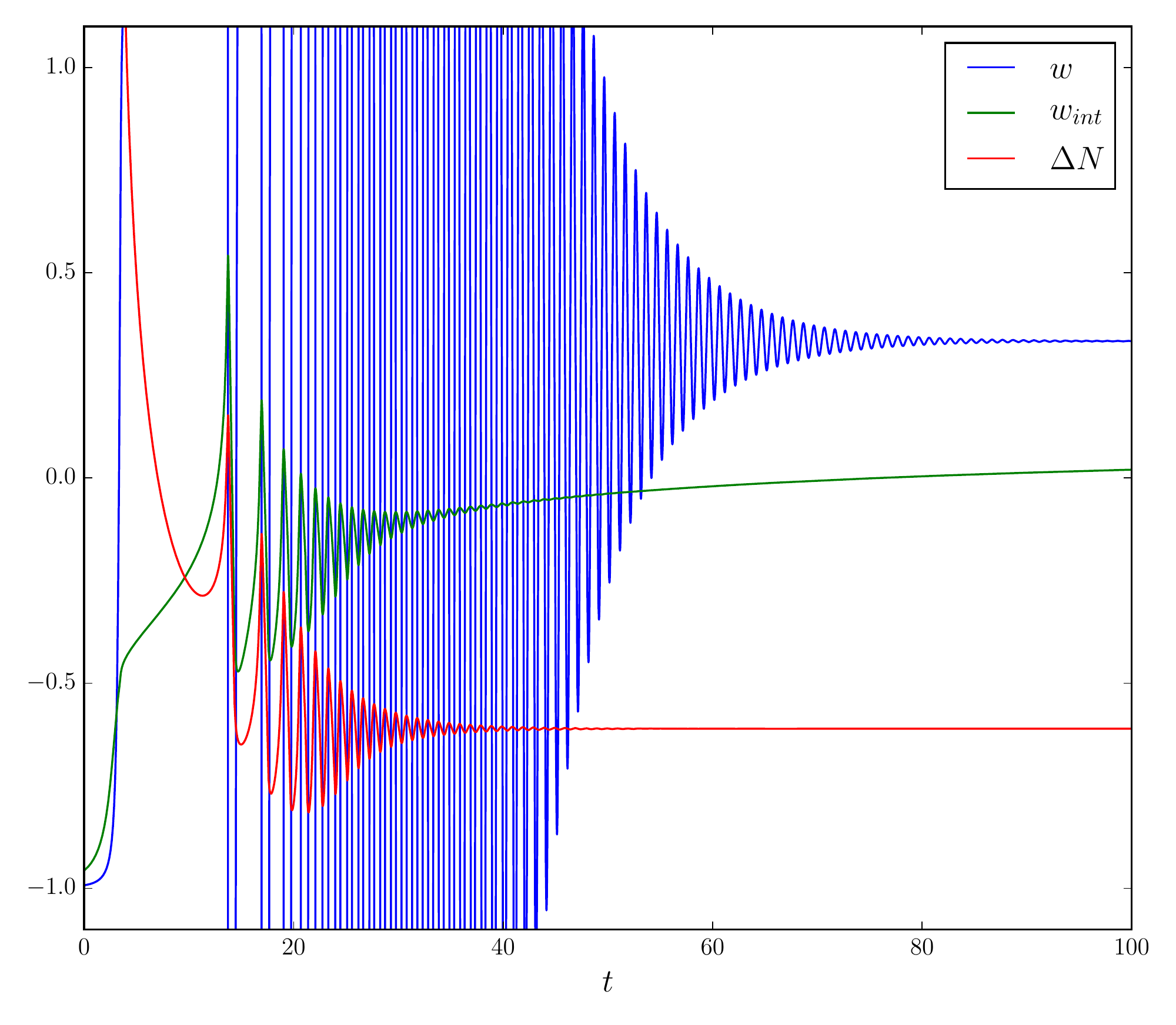}
    \includegraphics[width=0.5\textwidth]{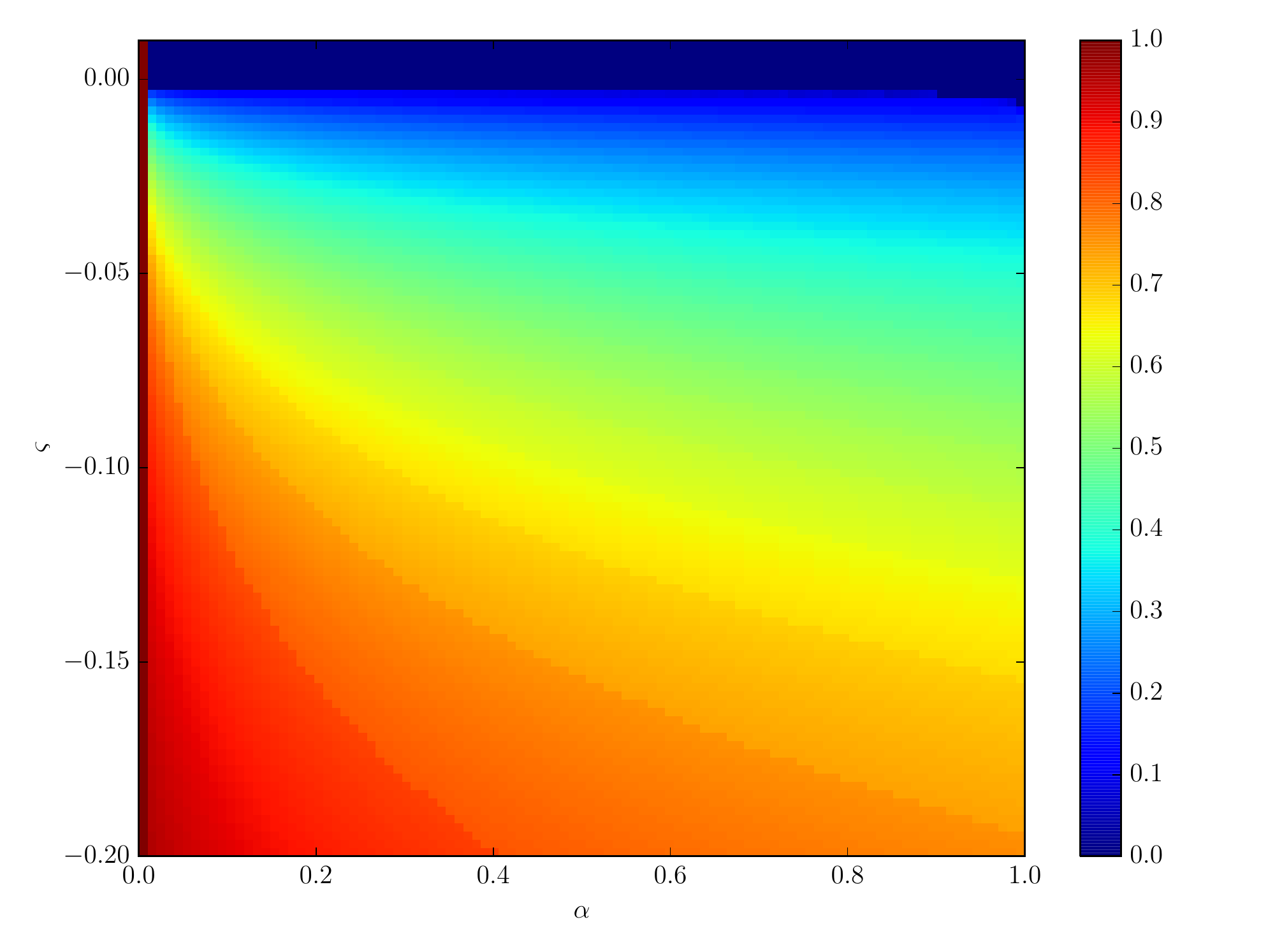}
    \caption{A typical evolution of the effective equation of state (left,blue), its e-fold average (green) and the resulting shift in the observable window $\Delta N$ (red) during reheating, as well as the reheating temperature as a function of Gauss-Bonnet coupling parameters $\alpha$ and $\varsigma$ (right). The equation of state oscillates wildly, even outside of the range $[-1,1]$ due to the non-standard coupling, but the physically relevant quantity derived from this, $\Delta N$, is fairly stable even at earlier times, justifying our method for defining the end of reheating. At late times the equation of state oscillates around and approaches $1/3$ as expected for successful reheating that gives way to a radiation-dominated epoch. We have used $n = m = 2$ and $\alpha = 0.1$, $\varsigma = -0.05$ in this example. In the right plot, temperature is scaled as a fraction of the $\alpha = 0$ (Gauss-Bonnet-free) case ($T_\text{r} \approx \mathcal{O}(10^{13} \si{\giga \electronvolt})$) such that a reading of $0.5$ corresponds to half the standard reheating temperature and so on. For small $\alpha$ and/or large negative $\varsigma$ the reheating temperature is barely changed, but for stronger Gauss-Bonnet effects the reheating temperature can decrease in principle down to zero, where reheating doesn't occur, such as for positive $\varsigma$.}
    \label{fig:reheatingf5}
\end{figure}

We define the end of reheating in our numerical simulations to be at the point when $\Omega_\gamma = 1 - x$, where $x$ is small \cite{Ellis:2015pla}. As the result for $w_\text{int}$ and hence $\Delta N$ we obtain will depend on our choice of $x$, we check that as time increases and $x$ approaches 0, that $\delta N$ approaches a fairly constant value such that our results are satisfactorily stable despite the free choice in defining when reheating has finished. This is shown in the left panel of figure \ref{fig:reheatingf5} for a typical choice of parameters. The value of $x$ decreases with time as radiation becomes more dominant. We see that while $w_\text{int}$ varies somewhat even at times moderately later than the end of inflation, $\Delta N$ is well described by small oscillations around a constant value once the post-inflationary oscillations settle down. From this we conclude that while we can't entirely remove the dependence of our numerical results on $x$, the precise value of $x$ chosen, if it is somewhat small, should not affect our conclusions drastically due to the reasonable stability of $\Delta N$, which is the quantity which directly affects the physics through eq. (\ref{eq:krange_phys}). The results presented here use $x = 1/3$, at which point the energy density of radiation is double that of the inflaton. We also ran simulations for smaller values of $x$ at which radiation is more strongly dominant, however we found numerically that there is very little dependence of the results on this choice, with $x = 0.1$ or $0.05$ giving largely indistinguishable results to the ones presented here, confirming our expectations that this should be the case from the near-constant nature of $\Delta N$ shown in \ref{fig:reheatingf5}.

The reheating temperature, $T_\text{r}$, may also be computed via the relation,

\beq \label{eq:reheattemperaturedef}
\rho_\text{th} = \frac{\pi^2}{30} g_{*} T_\text{r}^4 \, ,
\eeq
which again depends on the energy density at the end of reheating, $\rho_\text{th}$, as well as on the number of relativistic degrees of freedom $g_\text{th}$. The value of $T_\text{r}$ for a range of $\alpha$ and $\varsigma$ values in the intervals $[0,1]$ and $[-0.2,0]$ for $\Gamma = 4 \times 10^8$  is shown in the right panel of figure \ref{fig:reheatingf5}, as a convenient measure of how much the presence of the inverse Gauss-Bonnet coupling impedes and hence cools the reheating process. As we expected from our qualitative discussion in the previous section, we see that the presence of a Gauss-Bonnet coupling reduces $T_\text{r}$ by impeding the transfer of energy from inflaton oscillations to radiation.

\begin{figure}[!ht]
    \centering
    \includegraphics[width=0.48\textwidth]{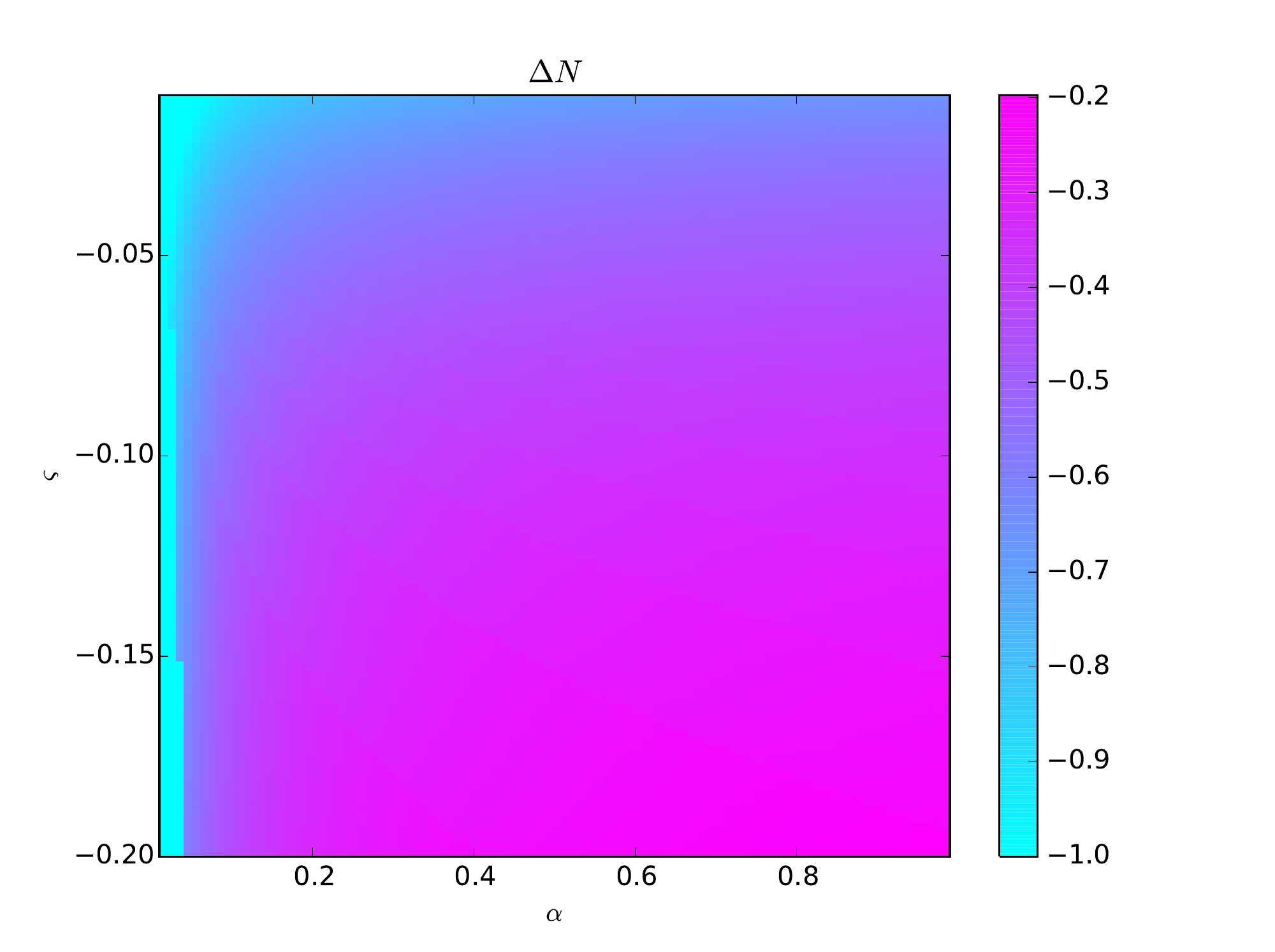}
    \includegraphics[width=0.48\textwidth]{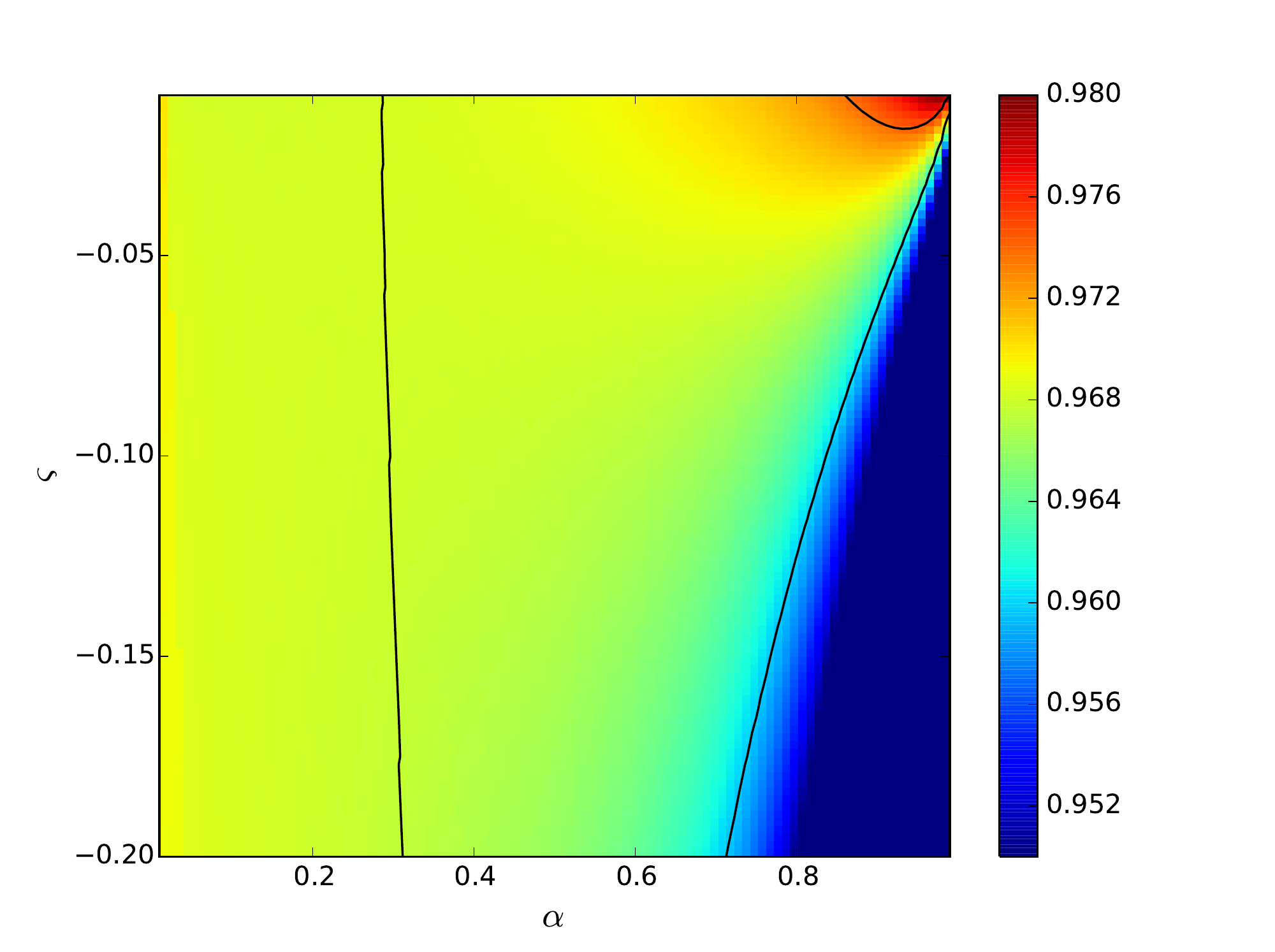}
    \caption{$\Delta N$ and the reheating-corrected scalar spectral index as a function of $\alpha$ and $\varsigma$. The left $\Delta N$ plot indicates the relative effect of reheating on the location of the observable window for different Gauss-Bonnet couplings. Contours have been drawn on the right $n_s$ plot indicating the areas disfavoured by current data. In particular, the region of parameter space to the left of the line at around $\alpha = 0.3$ is excluded as it predicts $r_{0.002} > 0.1$ and the regions on the right of the plot bound by the other contours have spectral indices excluded at $1\sigma$ by Planck. The remaining allowed region of parameter space is consistent with $A_s \approx 2.18 \times 10^{-9}$, and with scalar and tensor runnings of $\mathcal{O}(10^{-3})$ or less, in agreement with present CMB experiments.}
    \label{fig:reheatingf6}
\end{figure}

Finally, we calculate the primordial power spectra for the theory (as done in e.g. \cite{Guo:2010jr, PhysRevD.93.063519}) and evaluate them numerically at the appropriate Planck pivot scales using eq. (\ref{eq:krange_phys}) for the same range of parameters as above. Figure \ref{fig:reheatingf6} shows the scalar spectral index obtained by this procedure, along with the value of $\Delta N$ in each case as a measure of the effect of the variable reheating dynamics.

\section{Conclusions} \label{sec:Conclusion}

We have presented a study on the feasibility of models of inflation with an inverse power law coupling to the Gauss-Bonnet combination, taking into account the late-time absence of oscillations, which had previously gone unnoticed, by going beyond the slow-roll approximation. Having confirmed that neither standard (p)reheating, nor more involved mechanisms such as instant preheating, are able to successfully end inflation and reheat the universe, we allowed a slight generalisation in the model in which a small shift in the inverse power law coupling function, $\varsigma$, weakens the effect of the Gauss-Bonnet term, permitting reheating to proceed. In more general terms, this means that couplings to the Gauss-Bonnet term which become large towards the end of inflation have the potential to impede the onset of post-inflationary cosmological evolution.

We found in this model that especially for smaller values of $\varsigma$, the dynamics of the reheating phase are significantly different to those of fiducial models of inflation, with irregular oscillations and non-standard average equation of state. To investigate this reliably, we checked the validity of our method of parametrising the end of reheating, and found an exact result for the average equation of state during inflation to avoid the problem of numerically integrating many different rapidly oscillating functions. Following this, we computed the variation in the position of the observable window during inflation due to the modified reheating dynamics and computed the primordial power spectra for a range of parameters to obtain constraints on the model to demonstrate how these effects on reheating could be tested. Note that in this modified model, not only the case of perturbative reheating studied in detail here, but (instant) preheating and other non-perturbative methods of reheating which function in standard inflation, are also expected to be able to proceed for suitable values of $\varsigma$. We did not consider these mechanisms in detail here as satisfactory and interesting results were obtained from simple perturbative reheating, fulfilling our goal of understanding whether these models of inflation are feasible at all. As with perturbative reheating, however, we expect that the Gauss-Bonnet coupling will similarly be able to impede and modify such other mechanisms if the time at which particle production occurs coincides with a time when the Gauss-Bonnet coupling is large, which may have interesting implications.

Our key finding is that in models of inflation with an inverse power law coupling to the Gauss-Bonnet term, or more generally models where the coupling becomes very large at late times, reheating does not proceed in the usual fashion, or at all, depending on the strength and nature of the coupling. While this effect is important at the end of inflation, and hence has little to no effect on predictions of the power spectra (at best, shifting the observable window and hence the spectral properties by a few percent) in these models, we note that Gauss-Bonnet couplings in inflation may still provide an interesting mechanism for a low tensor to scalar ratio and otherwise reasonable predictions, it is important to be aware of the implications such a coupling has on the feasibility of the model from the perspective of achieving realistic post-inflationary cosmologies. These results, while computed explicitly for a shifted inverse power law coupling, are expected to be generic in any theory where the Gauss-Bonnet coupling becomes very large at late times and opposes the evolution of the inflaton. Even if the coupling does not formally become infinite at some point, as is the case here for simplicity, we expect this to be a good prototypical example for such situations. 

Interesting avenues to explore in future work on this topic include the details of Gauss-Bonnet coupled dynamics of preheating and what implications this may have for problems such as baryogenesis, and models where an additional scalar field is present in inflation but not coupled to the Gauss-Bonnet term, a scenario which may allow reheating to proceed through decays of the second field. Here, the inability of the Gauss-Bonnet coupled field to significantly decay, presented in this work as a problem, would be turned into a potential strength of the model as it could allow the inflaton to also serve as dark energy, providing a realisation of so-called quintessential inflation. Assuming we could realise such a situation satisfactorily, the Gauss-Bonnet coupling strength parameter $\alpha$ would need to take on sufficiently large values (early analysis indicates $\alpha > \sim 0.72$) to make the late-time value of $\epsilon_0$ in the dark energy epoch sufficiently small that the equation of state of the dark energy is acceptably close to $-1$ to meet experimental constraints. The details of this however depend heavily on the nature of the second field and how and when it may make inflation end, making a full analysis of such a scenario beyond the scope of the present work. It is possible that only a small region of parameter space will be able to simultaneously satisfy constraints coming from both inflation and dark energy. 

Another question to be addressed is whether our results affect the conclusions drawn in \cite{Herranen:2015ima}. In that paper, the authors showed that a coupling of the Higgs field to the Ricci scalar $R$ causes parametric resonance due to an effective oscillatory mass provided by the coupling. In the case studied in this paper, the evolution of the effective equation of state, and hence $R$, is non-standard and it would be worthwhile to investigate whether the Gauss--Bonnet coupling exaggerates or mitigates the production of Higgs bosons at the end of inflation.

\begin{acknowledgments}
The authors thank Mat Robinson for his contributions in the preliminary stages of this project. The work of CvdB and KD is supported by the Lancaster--Manchester--Sheffield Consortium for Fundamental Physics under STFC Grant No. ST/L000520/1. CL is supported by a STFC studentship. 
\end{acknowledgments}

\appendix

\section{Derivation of the analytical result for $w_\text{int}$}

In this appendix we seek to derive eq. (\ref{eq:wintexact}), an analytical result for the e-fold averaged equation of state during reheating. While we have not found a similar technique used elsewhere in the literature, we feel it is a prudent approach to the calculation as it avoids the problems of numerically integrating a rapidly oscillating function, and few assumptions are needed.

To begin, we assume the Friedman equations can be written in the form,

\begin{align}
3 H^2 & = \rho_\text{eff} \, , \label{eq:efff1}\\
2 \dot{H} & = -(\rho_\text{eff} + p_\text{eff}) \label{eq:efff2} \, .
\end{align}

The effective  equation of state is then $w_\text{eff} = p_\text{eff} / \rho_\text{eff}$. We then note that by dividing eq. (\ref{eq:efff2}) by eq. (\ref{eq:efff1}), we obtain,
 
\beq
\epsilon_0 = -\frac{\dot{H}}{H^2} = \frac{3}{2} (1 + w_\text{eff})  \quad \Rightarrow \quad w_\text{eff} = \frac{2}{3}\epsilon_0 - 1 \, .
\eeq
 
We then seek to use this result to evaluate the integral (\ref{eq:wintdef}), and find that,

\beq 
w_\text{int} = \frac{1}{N_\text{th} - N_\text{end}} \int^{N_\text{th}}_{N_\text{end}} \rpar{\frac{2}{3}\epsilon_0 - 1} \, \bd N \, .
\eeq

Using $N$ as a time coordinate (with $dN = H dt$), we have $\epsilon_0 = -H' / H$, so,

\beq
w_\text{int} = - \frac{2}{3} \frac{1}{N_\text{th} - N_\text{end}} \int^{N_\text{th}}_{N_\text{end}}\frac {H'}{H}  \, \bd N - \frac{1}{N_\text{th} - N_\text{end}} \int^{N_\text{th}}_{N_\text{end}}  \, \bd N \, .
\eeq

Having now transformed the integrand into something of the form $f'/f$ we can use standard results to evaluate this as,

\beq
w_\text{int} =  \frac{2}{3} \frac{\ln\rpar{H_\text{end} / H_\text{th}}}{\ln\rpar{a_\text{th} / a_\text{end}}} - 1 \, ,
\eeq

where we have used $N = \ln\rpar{a}$ and $-\ln\rpar{x} = \ln\rpar{x^{-1}}$ to simplify. Hence we have obtained the result (\ref{eq:wintexact}). As a consistency check, we confirm this is consistent with the usual result for single-fluid dominated cosmologies with constant equations of state, for which $a \propto t^{2/3(1+w)}$ and $H = 2/3(1+w)t$ and in the special case of $w = -1$ for which $H = \text{const}$ also correctly produces the correct result. Numerically, we further observe that for realistic early-time cosmologies with even rapidly varying equations of state, there is good agreement between this analytical result and the direct numerical integration of eq.  (\ref{eq:wintdef}).

\bibliography{GBreheatingrefs}

\end{document}